\documentstyle[12pt]{article}
\begin{document}
\newcommand{\newc}{\newcommand}

\newc{\be}{\begin{equation}}
\newc{\ee}{\end{equation}}
\newc{\ba}{\begin{eqnarray}}
\newc{\ea}{\end{eqnarray}}
\newc{\ie}{{\it i.e.} }
\newc{\eg}{{\it e.g.} }
\newc{\etc}{{\it etc.} }
\newc{\etal}{{\it et al.} }
\newc{\ra}{\rightarrow}
\newc{\lra}{\leftrightarrow}
\newc{\no}{Nielsen-Olesen }
\newc{\lsim}{\buildrel{<}\over{\sim}}
\newc{\gsim}{\buildrel{>}\over{\sim}}

\begin{titlepage}
\begin{center}
March 1994\hfill
             \hfill
\vskip 1in

{\large \bf
Stable, Spinning Embedded Vortices
}

\vskip .4in
{\large Leandros Perivolaropoulos}\footnote{E-mail address:
leandros@cfata3.harvard.edu},\footnote{Also, Visiting Scientist,
Department of Physics, Brown University, Providence, R.I. 02912.}\\[.15in]

{\em Division of Theoretical Astrophysics\\
Harvard-Smithsonian Center for Astrophysics\\
60 Garden St.\\
Cambridge, Mass. 02138, USA.}
\end{center}
\vskip .2in
\begin{abstract}
\noindent

The spinning vortex is a stationary generalization of the Nielsen-Olesen vortex
involving a
linear time dependence of the Goldstone boson. Here we show that this vortex
can be embedded in models with $SU(2)_{global}\times U(1)_{local}$ symmetry.
We also map the stability sector in parameter space and show
that for sufficiently large spinning velocities the vortex is stable for
{\it any} value of the Higgs self coupling parameter $\beta$.
This is a significant improvement of stability
compared to the semilocal vortex introduced by
Vachaspati and Achucarro which is stable only for $\beta<1$.
This result may have significant implications for electroweak vortices.
\end{abstract}

\end{titlepage}

\section{Introduction}
Topological defects form naturally in systems exhibiting spontaneous symmetry
breaking. The universality of the concept of symmetry breaking makes
topological defects fairly generic objects forming in both condensed matter and
particle physics theories.

The \no vortex\cite{no73}
is a particularly interesting
type of topological defect.
It plays a crucial role in cosmology\cite{v85} (gauged and
global cosmic strings), in particle physics (electroweak strings\cite{v92},
grand
unification strings\cite{ot82}) and in condensed matter
(superconductivity\cite{gl50,a57},
superfluids\cite{ds89,gkpb90},
liquid crystals\cite{cty91}).

Topological \no vortices normally form in systems where the first homotopy
group
of the vacuum manifold $M$ is nontrivial ($\pi_1 (M)\neq 1$). However, it was
recently shown\cite{av91,v92,vb92} that this is a sufficient but not necessary
condition
for their
formation. Vortices, dynamically stable to small perturbations may
even form in the standard electroweak model for a certain range of parameters.
Unfortunatelly, this range does not include the physically realized
parameter values\cite{jpv92}. Therefore, the following question arises:
{\it Can we find vortices that have better stability properties when embedded
in theories with trivial $\pi_1$ (like the standard electroweak model)?}

Previous attempts to address this question\cite{vw93} have shown that {\it
bound states} may
indeed improve the stability of embedded vortices but it was not clear if this
improvement was enough to stabilize the electroweak vortex for the physically
realized
values of parameters. An alternative approach involves a generalization of  the
\no vortex and a subsequent embedding in extensions of the electroweak model
\cite{l93,p93a,ej93,ds93}
(\eg the two doublet model). This approach has led\cite{ds93} to an interesting
new  stable embedded vortex involving a combination of two Higgs doublets. The
stability of
this vortex is due to the presence of an extra global symmetry.

Here we follow a different approach. We consider a single scalar doublet
and vortices with no bound states
but with a linearly time dependent Goldstone boson (spinning vortices). In
order to
achieve this spinning state, the existence of a charged background is
necessary. Such a background could be obtained for example by coupling to an
external charged
field with coherence length much larger than the width of the vortex, by a
background of
ions\cite{d90} or by a superfluid phase transition in the early
universe\cite{dw90,g91}.
Topologically stable spinning vortices have been shown to provide a description
of the
macroscopic properties of flux tubes in superconductors\cite{d90}.

 Embedded spinning vortices have some
similarities to embedded Cherns-Simons charged vortices\cite{pk86,k92} but
the fields of the former have different asymptotic behavior leading to
significantly improved stability properties. Another interesting study of
embedded charged
vortices can be found in Ref. \cite{a92} where a different ansatz from the one
discussed here was used to show that a charged vortex can be stable in a theory
with trivial
$\pi_1$.

Here, we show that spinning vortices, embedded in theories with trivial $\pi_1$
have significantly better stability properties compared to the corresponding
\no
embedded vortices (semilocal strings\cite{av91,p92}).

\section{The Embedded Spinning Vortex}

Consider the $SU(2)_{global}\times U(1)_{local}$ symmetric Lagrangian density
\be
{\cal L}=-{1\over 4} F_{\mu \nu}F^{\mu \nu} +{1\over 2} \vert D_\mu \Phi \vert
^2
-{\lambda \over 4} (\vert \Phi \vert ^2 - \eta ^2)^2 - A_\mu J^\mu
\ee
where $\Phi$ is a complex doublet scalar,
$D_\mu = \partial_\mu - ie A_\mu$, $F_{\mu \nu}=\partial_\mu A_\nu-
\partial_\nu A_\nu$ and $J_\mu=(\rho ({\vec x}), 0, 0, 0)$ represents a
charged background density with $\rho ({\vec x})\rightarrow \rho_0$ as
$r\rightarrow \infty$.

  Consider now the ansatz
\be
\Phi=\left( \begin{array}{c}
0\\
f(r) e^{i m \theta} e^{i\omega_0 t}
\end{array} \right)
\ee
\be
A_\theta = {{v(r)}\over r}, \hspace{1cm} A_0 = \alpha (r)
\ee
The difference between the embedded \no semilocal ansatz\cite{av91} and the one
of
(2), (3) is the time dependent phase $e^{i \omega_o t}$ which introduces a
linear time dependence in the Goldstone boson. We have also allowed for the
possibility of having $A_0 \neq 0$.

Choosing the gauge
\be
\lim_{r\rightarrow \infty} A_0 = 0, \hspace{1cm}
\partial_\mu A^\mu = 0
\ee
we obtain the equations of motion for $f(r)$, $v(r)$ and $\alpha (r)$
as follows

\be
f'' +  {{f'}\over r} + (\omega_0-\alpha)^2 f - {{(m-v)^2}\over r^2} f -\beta
(f^2-1) f=
0
\ee
\be
v'' - {{v'}\over r} + 2\hspace{1mm} f^2 (m-v)=0
\ee
\be
\alpha '' +{{\alpha '}\over r}+ 2\hspace{1mm} f^2 (\omega_0-\alpha)=-2 \rho (r)
\ee
where $'$ denotes derivative with respect to the azimuthal radius $r$, we have
rescaled
$f\rightarrow \eta\hspace{1mm} f$, $r\rightarrow  \sqrt{2}\hspace{1mm}
 r/(\eta\hspace{1mm} e)$, $v\rightarrow v/e$, $\omega_0 \rightarrow
\hspace{1mm} e \eta \hspace{1mm}\omega_0/\sqrt{2}$, $\alpha \rightarrow \eta
\hspace{1mm} \alpha /\sqrt{2}$, $\rho (r) \rightarrow (e^2 \hspace{1mm}
\eta^3/\sqrt{2})
\hspace{1mm} \rho(r)$ and defined $\beta \equiv 2\lambda/e^2$.

 The boundary
conditions to be imposed on $f$, $v$ and $\alpha$ in order to solve the system
of (5), (6) and
(7) are
\ba
r\rightarrow 0 &\Rightarrow & f\rightarrow 0, \hspace{3mm} v\rightarrow
0,\hspace{3mm} \alpha
' \rightarrow 0 \\
 r\rightarrow \infty &\Rightarrow & f'\rightarrow  0, \hspace{3mm}
v\rightarrow m, \hspace{3mm} \alpha \rightarrow 0
\ea
 These boundary conditions automatically fix the value of $\omega_0$ given the
value of
the parameter $\rho_0$ (see equation (23) below).

The $U(1)$ charge $Q_v$ of the configuration (2), (3) may be obtained
from the current $J^\mu_v ={1\over 2} i ((D^\mu \Phi)^\dagger \Phi
- \Phi^\dagger D^\mu \Phi)$. It is easily shown that
\be
Q_v=\int_V d^2 x J^0 =\int_V d^2 x\hspace{1mm} f^2 \hspace{1mm}
(\omega_0-\alpha) \simeq f^2_\infty
\hspace{1mm} \omega_0 V \ee
where $f_\infty$ is the asymptotic value of $f$ and $V$ is a large cutoff
volume in two dimensions
expressing the range of the background.  The angular momentum ${\vec M}_v$ of
the configuration may
also be found as
\be
{\vec M_v}=\int_V d^2 x \hspace{1mm}{\vec r} \times ({\vec E_v}\times {\vec
B})=-2\hspace{1mm} \int_V
d^2 x
\hspace{1mm}
f(r)^2 \hspace{1mm} (\omega_0-\alpha) \hspace{1mm} v(r) \hspace{1mm} {\hat e}_z
\simeq -2
\hspace{1mm}  m\hspace{1mm}Q_v \hspace{1mm} {\hat e}_z
\ee
where $E_v$ is the electric field induced by the charge of the vortex.
The charge $Q_{bg}$ and angular momentum ${\vec M}_{bg}$ of the background
charge density may
also be obtained and shown to cancel the divergences of the corresponding
vortex quantities
\be
Q_{bg}=\int_V d^2 x \hspace{1mm} \rho (r) \simeq \rho_0 V \hspace{1mm} \simeq
-Q_v
\ee
\be
{\vec M}_{bg}=\int_V d^2 x \hspace{1mm} {\vec r} \times ({\vec E_{bg}}\times
{\vec B})
=-2\hspace{1mm} \int_V d^2 x \hspace{1mm}
\rho (r)\hspace{1mm} v(r) \hspace{1mm}{\hat e}_z \simeq
-2\hspace{1mm}m \hspace{1mm} Q_{bg} \hspace{1mm} {\hat e}_z
\ee
where $E_{bg}$ is the electric field due to the background
charge density $\rho (r)$.
The total charge $Q_{tot}=Q_v + Q_{bg}$ and total angular momentum ${\vec
M}_{tot} ={\vec M}_v + {\vec M}_{bg}$ are in general non-zero, finite and
conserved.

The connection between background charge and vortex angular momentum shown in
(10)-(13) implies that
injecting charge into the vacuum in the presence of strings is equivalent to
spinning up
the vortices.
This property has been used
previously\cite{g91} to construct a cosmological model where large scale
coherent velocity fields are generated {\it non-gravitationally} by global
topological
vortices
which spin up due to the injection of charge provided by a temporary late time
superfluid
phase.

\section{Stability}

Consider now the following small perturbation to the ansatz (2), (3)
\be
\delta \Phi  =
\left(
\begin{array}{c}
  g(r) e^{i n \theta} \\
\delta f (r, \theta )
\end{array}
\right)
\ee
\be
\delta A_\mu (r,\theta)
\ee
The energy of the perturbed configuration decouples as follows
\be
E= E_0 (f,v,\alpha) + \delta E (\delta f, \delta A_\mu) + E_1 (g)
\ee
where
\be
E_0 = \int_V d^2 x \hspace{1mm} [f'^2 +\alpha'^2 +{{v'^2}\over
{2\hspace{1mm}r^2}} +
(\omega_0 -\alpha)^2 f^2  +
{{(m-v)^2}\over {r^2}} f^2 +  {\beta \over 2}
(f^2-1)^2 + \rho \hspace{1mm} \alpha]
\ee
is the energy of the unperturbed solution which is identical to
the energy of the {\it topologically stable} charged spinning vortex
obtained by substituting the complex doublet in the ansatz (2) by a complex
singlet.
 The term $\delta E (\delta f, \delta A_\mu)$ is identical to the one obtained
by
perturbing the topologically stable charged spinning vortex and therefore it is
non-negative.
 Thus, the term that determines the stability of the embedded
spinning vortex is $E_1 (g)$ which to second order in $g$ is
\be
E_1 = \int_V d^2 x \hspace{1mm} [g'^2 + \alpha^2 g^2 + {{(n-v)^2}\over {r^2}}
g^2 +
\beta (f^2 -1) g^2]
\ee
 Stability is obtained
for that range of parameters $(\rho_0, \beta)$ for which $E_1 (g)$ is positive
definite.
Stability also depends on the detailed functional form of $\rho (r)$ around the
core of the
vortex which in turn can be obtained by specifying the dynamics of the
background charge
density. In order to keep our analysis simple and general we will not assign
any special
dynamics to $\rho (r)$. Instead we will study two opposite extreme forms of it.
Most
dynamically obtained forms of $\rho (r)$ are expected to be between these two
extreme cases.

The first extreme is that of a {\it
soft} background which completely adjusts itself to the shape of the core thus
neutralizing the charge density induced by the spin of the vortex. For such a
background
\be
\omega_0 f^2 (r) = -\rho (r), \hspace{1cm} \alpha=0
\ee

The second extreme is that of a {\it hard} background which is not affected by
the presence of
the vortex. In this case
\be
\rho (r) = \rho_0, \hspace{1cm} \alpha \neq 0
\ee
 everywhere within the range of the background.

We first study the stability in a soft background. In this case $E_1$ is
non-negative
$iff$ the following eigenvalue equation has only non-negative eigenvalues
$\omega^2$
\be
-g''- {{g'}\over r} + {(v-n)^2 \over r^2} g + \beta (f^2-1)g  = \omega^2 g
\ee
 In what follows we consider $n=0$ in order to examine eigenstates of lowest
energy.
Notice that both parameters $\rho_0$ and $\beta$ enter in (21) since the
unperturbed fields $f$,
and $v$ depend on $\rho_0$. In particular $\rho_0$ determines the spinning
velocity
$\omega_0$ which in turn specifies the asymptotic behavior of $f$ as
\be
\lim_{r\rightarrow \infty} f^2 = (1+ {{\omega_0^2}\over \beta})\equiv \gamma^2
\ee
where from (19) $\omega_0$ is the solution of
\be
\gamma^2 \omega_0 = - \rho_0
\ee

We have solved the system of (5), (6) ((7) is trivial for a soft background)
with boundary conditions (8), (9), using a relaxation scheme based on Gaussian
elimination\cite{nr}.  Fig. 1 shows the potential of the Schroedinger-like
eigenvalue equation (21)
(dotted line) with $\beta=15$, $\rho_0=2$, compared with the corresponding
potential (continous line)
obtained in the case of an embedded {\it non-spinning} vortex (semilocal
string, $\beta=15$,
$\rho_0 =0$). Clearly the potential in the case of the spinning vortex is more
positive and
thus less receptive to negative eigenvalues.

The improvement of stability may also be seen by a rescaling $r\rightarrow
r/\gamma$,
$f\rightarrow \gamma f$ which transforms (5) and (6) (with $\alpha=0$) to
exactly the \no equations.
The potential of the Schroedinger-like equation (21) becomes
\be
V(r) = {v^2 \over r^2} + \beta (f^2-1/\gamma^2)
\ee
The case of the non-spinning (semilocal) vortex is obtained in the limit
$\gamma \rightarrow 1$ (\ie  $\omega_0 \rightarrow 0$). In this limit the
stability problem has been solved by Hindmarsh\cite{h92}
(see also \cite{akpv92,l92}) who found that there are no
negative eigenvalues for $0\leq \beta \leq 1$ (implying stability) while for
$\beta > 1$ there are negative eigenvalues and the embedded vortex is unstable.
This instability is the main reason that the embedded electroweak vortex is
unstable
for the physically realized values of parameters of the standard electroweak
model\cite{jpv92}.

In the case of the spinning vortex we have $\omega_0 \neq 0$ or $\gamma^2 > 1$
and the potential (24) is less receptive to negative
eigenvalues. In fact, as $|\omega_0 |\rightarrow \infty$ (\ie
$|\rho_0|\rightarrow
\infty$), $V(r)$ is positive definite and the vortex is stable for {\it all}
values of $\beta$.

 The obtained solution for $f$, $v$ with winding number unity\footnote{We have
also considered
higher values of
winding number $m$ and shown that the stability does not improve in those
cases}
($m=1$) was used to solve the eigenvalue equation (21).
A fourth order
Runge-Kutta algorithm\cite{nr} was implemented for the solution.

The points on the continous line in Fig. 2 represent critical values of
$\rho_0$ (for
fixed $\beta$) such that for $\rho_0 > (\rho_0)_{crit}$ there are no
negative eigenvalues, while for $\rho_0 < (\rho_0)_{crit}$ we found negative
eigenvalues. Thus sector I in Fig. 2 is the sector of stability in a soft
background while for
parameter values in sectors II and III the embedded spinning vortex is
unstable.
In the
limit $\rho_0 \rightarrow 0$ we recover the semilocal vortex case (stability
for $\beta <1$, instability for $\beta > 1$) as expected. Also, sector I is
consistent with the expected stability for all positive values of $\beta$ in
the limit $\rho_0 \rightarrow \infty$.

We now consider the opposite extreme background form: {\it the hard
background}. In this case
$\alpha \neq 0$ and the eigenvalue equation (21) becomes
\be
-g''- {{g'}\over r} + {(v-n)^2 \over r^2} g + \beta (f^2-1)g  + \alpha^2 g=
\omega^2 g
\ee
The potential of (25) has an additional positive contribution by the term
$\alpha^2$ which
may lead to the immediate conclusion that stability is further enhanced in this
case.
This additional positive definite contribution is obtained for {\it any} form
of background
density $\rho (r)$ which is different from the soft background form defined in
equation (19).
Any such background will excite the $A_0$ component of the gauge field and will
in general {\it
further improve} the stability of the embedded spinning vortex.
In order to test this conjecture we have explicitly solved the eigenvalue
equation (25) for the
opposite extreme case of the hard background defined in equation (20).
   Using a similar method as in the case of the soft background we have solved
the
unperturbed system (5), (6), (7) and used the solution to study (25). The
stability sector is
shown in Fig. 2 as the union of sectors I and II. In the parameter sector III
we found
negative eigenvalues which imply that the embedded spinning vortex with a hard
background is
unstable for parameters in this sector. It is clear from Fig. 2 that as
expected, a hard background
 improves further the stability of the spinning vortex. This may also be seen
in Fig. 1
where we plot the potentials for the eigenvalue equation (25) in the cases of
semilocal (continous
line), spinning in soft background (dotted line) and spinning in hard
background (dashed line)
vortices.

Clearly, the stability has improved dramatically compared to the non-spinning
embedded vortex
(semilocal string) in both types of background considered. Since the background
types
 are opposite extremes of each other we are led to the conclusion that a
background charge density inducing spin is a generically effective way to
stabilize the embedded vortex.

A simplified way to understand this enhancement of stability for the spinning
vortex is that the centrifugal force introduced by the rotation of the
Goldstone boson tends to prevent the tilt of the field $\Phi$ towards the
upper component thus stabilizing the vortex.

\section{Conclusion}

In conclusion we have shown that it is always possible to stabilize a vortex
embedded in an $SU(2)_{global}\times U(1)_{local}$ theory by introducing the
appropriate amount of angular momentum in the Goldstone boson. This angular
momentum may
be introduced by a charged background
density $\rho (r)$ which is also needed for long range neutrality. The role of
$\rho (r)$ can be played either by ions (in condensed matter systems) or by
a coupling to a dynamical charged field with coherence length much larger than
the width of the vortex.

Some exciting questions remain to be addressed:
\begin{itemize}
\item
What are the stability properties of the spinning vortex when embedded in the
standard electroweak model?
\item
Can embedded global vortices (\ie vortices involving no gauge fields) be
stabilized by the introduction of spin through a charged background?
\end{itemize}
Stable, spinning electroweak vortices could be effective in a non-gravitational
generation of large scale coherent velocities provided that a background charge
density
is generated by a late superfluid phase transition. Such a prospect
would give more realistic dimensions to the mechanism proposed in Ref.
\cite{g91}.

The second question is also particularly interesting in view of the fact that
non-spinning embedded global vortices are always unstable to
small perturbations\cite{bpv93}. Global spinning vortices are important in
correctly modeling
many features of condensed matter systems \cite{ds89,gkpb90} (\eg superfluids).
These issues are currently under investigation.

\bigskip

{\bf Acknowledgments}

\noindent
I would like to thank Robert Brandenberger, Manuel Barriola and Tanmay
Vachaspati for useful
comments on the manuscript.
 This work was supported by a CfA Postdoctoral
Fellowship.

\vspace{5mm}
\centerline{\large \bf Figure Captions}

{\bf Figure 1:}
The potential of the Schroedinger-like stability equation for
semilocal, spinning in soft background and spinning in hard background
vortices.
Notice that the potential favors negative eigenvalues less, when spin is
introduced.
\vskip 0.5cm
{\bf Figure 2:}
The stability map in parameter space.
The spinning vortex in a
soft background is stable for points in sector
I but unstable for points in sectors II and III.
In the case of a hard background stability occurs in both
sectors I and II while instability occurs in sector III.
For $\rho_0 = 0$ we recover the
semilocal string. $\rho_0$ is in units of $e^2\hspace{1mm}\eta^3 /\sqrt{2}$.


\vskip 1cm

\begin{thebibliography}{99}
\bibitem{no73} H. B. Nielsen and P. Olesen, {\it Nucl. Phys.} {\bf B61}, 45
(1973). \\
see also, L. Perivolaropoulos, {\it Phys. Rev.} {\bf D48}, 5961 (1993)
for a correction in the vortex asymptotic behavior presented by
Nielsen-Olesen and in most subsequent papers.
\bibitem{v85} A. Vilenkin, {\it Phys. Rep.} {\bf 121}, 265 (1985) and
references therein.
\bibitem{v92}  T. Vachaspati, {\it Phys. Rev. Lett.} {\bf 68}, 1977 (1992).\\
T. Vachaspati, {\it Nucl. Phys.} {\bf B397}, 648
(1993).
\bibitem{ot82} D. Olive and N. Turok, {\it Phys. Lett.} {\bf B117}, 193 (1982).
\bibitem{gl50} V. I. Ginzburg and L. D. Landau, {\it Zh. Eksp. Teor. Fiz.}
{\bf 20}, 1064 (1950).
\bibitem{a57} A. A. Abrikosov, {\it Zh. Eksp. Teor. Fiz.} {\bf 32}, 1442
(1957).\\
A. A. Abrikosov, {\it Sov. Phys. JETP} {\bf 5}, 1174 (1957).
\bibitem{ds89} R. Davis and P. Shellard, {\it Phys. Rev. Lett.} {\bf
63}, 2021 (1989).
\bibitem{gkpb90} B. Gradwohl, G. Kalbermann, T. Piran and E. Bertschinger,
{\it Nucl. Phys.} {\bf B338}, 371 (1990).
\bibitem{cty91}I. Chuang, N. Turok and B. Yurke, {\it Phys.Rev.Lett.}
{\bf 66}, 2472 (1991).
\bibitem{av91} T. Vachaspati and A. Achucarro, {\it Phys. Rev.} {\bf
D44}, 3067 (1991).
\bibitem{vb92} T. Vachaspati and M. Barriola, {\it Phys. Rev. Lett.} {\bf 69},
1867 (1992).
\bibitem{jpv92} M. James, L. Perivolaropoulos and T. Vachaspati,
{\it Phys. Rev.}{\bf D46}, 5232 (1992)\\
M. James, L. Perivolaropoulos and T. Vachaspati,
{\it Nucl. Phys.} {\bf B395}, 534 (1993).
\bibitem{vw93} T. Vachaspati and R. Watkins, {\it Phys. Lett.} {\bf B318}, 163
(1993).
\bibitem{l93} H. S. La, hep-ph/9302220 (1993).
\bibitem{p93a} L. Perivolaropoulos, {\it Phys. Lett.} {\bf B316}, 328 (1993).
\bibitem{ej93} M. Earnshaw and M. James, {\it Phys. Rev.} {\bf D48}, 5818
(1993).
\bibitem{ds93} G. Dvali and G. Senjanovic, {\it Phys. Rev. Lett.} {\bf 71},
2376 (1993).
\bibitem{d90} R. Davis, {\it Mod. Phys. Lett.} {\bf A5}, 955 (1990).
\bibitem{dw90} S. Dodelson and L. M. Widrow, {\it Phys. Rev. Lett.}
{\bf 64}, 340 (1990).\\
S. Dodelson and L. M. Widrow, {\it Phys. Rev.} {\bf D42}, 326 (1990).
\bibitem{g91} B. Gradwohl, {\it Phys. Rev.} {\bf D44}, 1685 (1991).
\bibitem{pk86} S. Paul and A. Khare, {\it Phys. Lett.} {\bf B174},
420 (1986).
\bibitem{k92} A. Khare, {\it Phys. Rev.} {\bf D46}, 2287  (1992).
\bibitem{a92} E. Abraham, {\it Nucl. Phys.} {\bf B399}, 197 (1992).
\bibitem{p92} J. Preskill, {\it Phys. Rev.} {\bf D46}, 4218 (1992).
\bibitem{nr} W. Press, B. Flannery, S. Teukolsky, W. Vetterling,
{\it Numerical Recipes, 2nd Edition},  Cambridge Univ. Press (1993).
\bibitem{h92} M. Hindmarsh, Phys. Rev. Lett. {\bf 68}, 1263 (1992)
\bibitem{akpv92} A. Ach\'ucarro, K. Kuijken, L. Perivolaropoulos
T. Vachaspati, {\it Nucl. Phys.} {\bf B388}, 435 (1992).
\bibitem{l92} R. Leese, {\it Phys. Rev.} {\bf D46}, 4218 (1992).
\bibitem{bpv93} M. Barriola, L. Perivolaropoulos and T. Vachaspati,
unpublished.
\end{thebibliography}
\end{document}